\documentclass[%
reprint,
 amsmath,amssymb,
 aps,
]{revtex4-2}

\usepackage{graphicx}
\usepackage{dcolumn}
\usepackage{bm}
\usepackage{braket}
\usepackage{lineno}
\usepackage{amsmath}
\usepackage{bm}
\usepackage{soul}
\usepackage{subcaption}
\usepackage{multirow,tabularx}
\usepackage{diagbox}
\usepackage{tabularx}
\usepackage[justification=raggedright, singlelinecheck=false]{caption}
\usepackage{array}
\usepackage{booktabs}
\let\vec\mathbf

\begin{document}

\preprint{APS/123-QED}

\title{Imaging Spinor Bose Gasses Using Off-Axis Holography}

\author{Nejc Blaznik}
 \email{n.blaznik@uu.nl}
\author{Jasper Smits}
 \altaffiliation[Also at ]{Tectonics, Department of Earth Sciences, Faculty of Geosciences, Utrecht University.}
\author{Marc Duran Gutierrez}
\author{Peter van der Straten }
\affiliation{%
Debye Institute for Nanomaterials Science \\and Center for Extreme Matter and Emergent Phenomena, \\Utrecht University}%

\date{\today}

\begin{abstract}
We introduce a novel, non-invasive imaging technique based on spin-dependent off-axis holography (SOAH) for spin-1 Bose-Einstein condensates (BECs). Utilizing a dual reference beam strategy, this method records two orthogonal circular polarization components of a single probe beam. The circular birefringence of spin-polarized atoms induces differing complex phase shifts in the polarization components of the light, which are reconstructed from the interference patterns captured on camera. Our approach enables spin- and density-resolved imaging of both phase and amplitude information \emph{in-situ} on a sub-millisecond time scale with minimal disturbance to the condensate. We explore the technique's efficacy under various background static fields, demonstrating its sensitivity to the quantization axis of the atoms and confirming its effectiveness.
\end{abstract}

\maketitle

\section{Introduction}
Spinor Bose-Einstein condensates (BECs) represent a quantum state of matter with vast potential for advancing our understanding of the universe's foundational principles. These systems, characterized by their multiple internal spin states lie at the interface between superfluidity and magnetic ordering. They present a unique platform for investigating phenomena not present in usual single-component BEC's, where the spin degree of freedom is frozen. This opens up possibilities to study various phenomena such as quantum phase coherence and transitions \cite{Chang:2005, Sadler:2006, Bookjans:2011}, topological defects \cite{Khawaja:2001, Makela:2003, Sadler:2006}, and spin dynamics \cite{Kawaguchi:2012, Black:2007}. 

Despite significant progress in BEC research over the last three decades, existing methodologies for probing spinor BECs are either invasive and destructive or lack adequate temporal, spatial, or spin sensitivity to capture the full spin composition \emph{in-situ}, in real time. Traditionally, the most straightforward tool for probing spinor condensates is through Stern-Gerlach (SG) splitting, where a magnetic field gradient is applied on an expanding cloud of atoms, which causes spatial separation of different spin components before imaging. After imaging, the initial spin distribution of the cloud can be retroactively surmised. However, this technique has a number of limitations that can obstruct the observation of physics of spinor condensates. It is intrinsically destructive, making it unable to probe any dynamics on a single condensate and the need for a minimum time-of-flight blurs the spatial resolution of the condensate's structure. Nevertheless, this technique has been used to study domain formation and relaxation \cite{Stenger:1998, Miesner:1999, Isoshima:1999, Black:2007, Jimenez-Garcia:2019}, spinor dynamics and phase transitions in antiferromagnetic spin-1 condensates \cite{Black:2007, Bookjans:2011, Vinit:2013}, and investigate magnetic solitons \cite{Bersano:2018, Chai:2019}. 

To mitigate atomic perturbations without significantly compromising spatial resolution, dispersive imaging techniques employing off-resonant light have been employed. The phase delay accrued by probe light as it traverses the sample serves as a direct indicator of the atomic cloud's refractive index. This effect is interpretable by converting the phase delay into intensity variations, captured on a camera through techniques such as phase-contrast imaging \cite{Andrews:1996, Meppeling:2009}, intensity-based defocus-contract imaging~\cite{Turner:04}, dark-field Faraday rotation imaging \cite{Gajdacz:2013}, shadowgraph imaging \cite{Wigley:2016} and off-axis holography \cite{Smits:2020}. Additionally, the interaction between the light and the condensate reveals intricacies beyond mere density distribution. Through the spin-dependent dispersive birefringence of the atoms, one is able to discern the different spin components through Faraday rotation. Such optical detection schemes have been employed to measure the magnetization in $F=1$ spinor gases of $^{87}$Rb \cite{higbie:2005, Sadler:2006, vengalattore2010periodic, Kaminski2012}, and of $^{23}$Na \cite{Liu:2009, SangWonSeo:2015}. The techniques used in rubidium do not translate well to sodium, due to the smaller hyperfine splitting in sodium. 
In sodium, the proximity of other hyperfine levels complicates this approach, potentially averaging out the spin dependence and limiting the techniques' effectiveness in dynamic studies when high spatial and temporal resolution are necessary. Furthermore, while these methods are effective in magnetization measurements, they do not allow for comprehensive spin density reconstruction.

In this paper, we introduce a novel non-invasive imaging technique based on spin-dependent off-axis holograhy (SOAH), offering several improvements over existing methods. A key feature of SOAH is the employment of a dual-reference beam strategy for independent reconstruction of both orthogonal polarization components of the probe beam. Since the full field information is obtained, each of the two components can undergo numerical correction for image defocusing, coma, and spherical aberrations \cite{Thylen1983, Smits:2020}. The two polarizations interact differently with different spin states of the condensate, thus allowing for \emph{in-situ} measurement of spin-dependent magnetization and density. Leveraging SOAH's inherent heterodyne gain, imaging is minimally destructive, permitting the collection of several hundreds of images per sample, which allows for the study of spin dynamics on a sub-millisecond timescale. Moreover, since SOAH does not require precise alignment of optics in the Fourier plane, common setup constraints are eliminated, enhancing the method's applicability.

\section{Methods}
\subsection{Imaging}
As an electromagnetic wave traverses a cloud of spin-polarized atoms, its propagation is influenced by the atoms' collective response, encapsulated by the refractive index $\mathcal{N}_q$, which depends on the polarization $q$ of the probe. In quantum gases, the phase $\phi_q$ that a beam accumulates is related to $\mathcal{N}_q$,
\begin{equation}\label{eq:phi_integral}
    \phi_q(x, z) = k \int\Big(\mathcal{N}_q(x, y, z) - 1 \Big) dy,
\end{equation}
with $k = \frac{2 \pi}{\lambda}$, and $\lambda$ the wavelength of probe light and the integration is along the line-of-sight \cite{Meppeling:2009}. Refractive index is in turn intricately linked to the gas density $\rho_m$, where $m$ denotes the magnetic substate, via the polarizability $\alpha_{q,m}$, and under typical experimental conditions, where $|\rho_m \alpha_{q,m} / \varepsilon_0| \ll 1$, it can be expressed as
\begin{equation}
    \mathcal{N}_q \approx 1 + \sum_m \frac{\rho_m \alpha_{q,m}}{2 \varepsilon_0},
\end{equation}
which simplifies Eq. (\ref{eq:phi_integral}) to     
\begin{align}\label{Eq:PhaseDelay_Pol}
    \phi_q (x, z) =& \ k \sum_m \frac{\alpha_{q,m}}{2 \varepsilon_0} \int \rho_m(x, y, z) dy \nonumber \\
    =& \ k \sum_m \frac{\alpha_{q,m}}{2 \varepsilon_0} \rho_m^c(x, z). 
\end{align}
Here $\rho_m^c(x, z)$ is the column density resulting from the integration along the line of sight. The polarizability $\alpha_{q,m}$ depends on the detuning of light $\delta_e$, and is given by 
\begin{equation}
    \alpha_{q,m}(\delta) = i \frac{\varepsilon_0 c \sigma_{\lambda}}{\omega} \sum_{e}\frac{{C}_{g, e}^q}{1 - 2 i \delta_e/\Gamma}, 
\end{equation}
with $\sigma_{\lambda} = {3 \lambda^2}/{2 \pi}$ the scattering cross section, $\omega$ the probe frequency and $\Gamma$ the natural linewidth. Here, $C_{g,e}^q$ is the Clebsch-Gordan coefficient for the transition $g\rightarrow e$, which depends on the polarization $q$ of the probe. These coefficients are not the same for $\sigma_+$ ($q=+1$), $\sigma_-$ ($q=-1$) and $\pi$ ($q=0$) polarized light. In our case the sum of coefficients for each substate $m$ can be calculated using Wigner 3-j and 6-j symbols \cite{metcalf1999laser} and are for experimental combinations of $q$ and $m$ given in Table~\ref{tab:CGs}. Note that as the polarizability is a complex parameter, so is the phase $\phi_q$. The real part of $\phi_q$ yields the phase shift induced by the atoms, while the imaginary part is related to scattering. 

In our technique we will only probe the phase shift of the beam. In case the detuning is large compared to the splitting of the upper states, Eq.~(\ref{Eq:PhaseDelay_Pol}) reduces to
\begin{equation}\label{Eq:PhaseShiftTwoComponents}
    \phi'_q = \sum_m \sigma_{q,m} \rho_m^c,
\end{equation}
where the $m$-dependent scattering cross-section $\sigma_{q,m}$ is given by
\begin{equation}\label{Eq:M_Scattering_CS}
    \sigma_{q,m} = - \sigma_\lambda \frac{(\delta/\Gamma)}{1+(2\delta/\Gamma)^2} \sum_{e}{C}_{g, e}^q.
\end{equation}
Our technique exploits the $m$-dependence of the cross-section to detect the different spin components. Furthermore, by choosing the detuning large compared to the linewidth, we can suppress the scattering and thus make the method inherent nearly non-invasive.

\begin{table}
    \caption{The sum of Clebsch-Gordan coefficients for each magnetic substate $m$ driven by light with polarization $q$ for our particular transition, namely for Na-atoms in the $F$=1 state excited on the $^3$P$_{3/2}$ transition. Here $q=+1$ refers to $\sigma_+$, $q=-1$ to $\sigma_-$, and $q=0$ to $\pi$ polarized light. When light is polarized linearly, perpendicular to the axis of magnetic field, there is no spin-dependent contrast.}
    \label{tab:CGs}
    \begin{tabularx}{0.45\textwidth}{c|>{\centering\arraybackslash}X|>{\centering\arraybackslash}X|>{\centering\arraybackslash}X}
        \hline \hline
        \diagbox[dir=NW]{$m$}{$q$} & \textbf{-1} & \textbf{0} & \textbf{+1} \\ 
        \hline
        \textbf{-1} & 1/2 & 2/3 & 5/6 \\
        \textbf{0}  & 2/3 & 2/3 & 2/3 \\
        \textbf{+1} & 5/6 & 2/3 & 1/2 \\
        \hline \hline
    \end{tabularx}
\end{table}

In our experiment, we simultaneously measure the phase delays of both polarizations by imaging an atom cloud with a linearly polarized probe beam, oriented perpendicular to the magnetic field. In this frame, the atoms perceive the linearly polarized beam as a coherent superposition of two circular polarizations, $\vec{E}_{p, in}= E_{p, x} \bm{\hat{x}} \equiv \frac{E_{p, x}}{\sqrt{2}} (\bm{\hat{\sigma}}_+ + \bm{\hat{\sigma}}_-)$. This causes the atoms to interact differently with each circular component based on their spins, leading to two distinct spin-dependent phase shifts. The atoms' impact on the probe field is thus characterized by 
\begin{align}
    \vec{E}_{p, out} = \frac{E_{p, x}}{\sqrt{2}} (e^{-i\phi_+}\bm{\hat{\sigma}}_+ + e^{-i\phi_-}\bm{\hat{\sigma}}_-),
\end{align}
with $\phi_{q = \pm 1} \equiv \phi_{\pm} = \phi'_{\pm} + i \phi''_{\pm}$ determined by the spin-dependent phase delay $\phi'$ and optical density $2\phi''$. After the beam exits the atom cloud, a quarter-wave plate converts the circular components to linear, allowing them to interfere with two linearly polarized reference beams. Calibration without the quarter-wave plate ensures alignment of all linear polarizations. For the rest of the treatment, we assume, without loss of generality, that the two reference beams are described by $\vec{E}_{R1} = E_{R1} \bm{\hat{x}}$, and $\vec{E}_{R2} = E_{R2} \bm{\hat{y}}$. Note that the two reference beams hit the detector under different angles with respect to each other, such that their interference patterns can be separated in Fourier space.

Analogous to the analysis in \cite{Smits:2020}, the intensity recorded on the camera, is an interference pattern of the probe and the two reference beam fields,
\begin{align}
    I \propto &|E_p|^2 +  |E_{R1}|^2 +  |E_{R2}|^2 \nonumber \\
    & + E_p E_{R1}^*    e^{-i (\phi_+ + \tilde{\bm{k}}_{p,R1} \cdot \bm{r})} \nonumber\\
    & + E_p E_{R2}^*    e^{-i (\phi_- + \tilde{\bm{k}}_{p, R2} \cdot \bm{r})} \\
    & + E_{R1} E_{R2}^* e^{-i (\tilde{\bm{k}}_{R1, R2} \cdot \bm{r})}  \nonumber \\
    & + c. c. \nonumber
\end{align}
where $\bm{r} = (x, z)$, $\tilde{\bm{k}}_{i, j} = \bm{k}_i - \bm{k}_j$ with $i,j=p, R1, R2$ is the difference wavevector of two incoming fields determined by the angle between the two beams, and $\phi_\ell$ with $\ell=\pm$ is the phase factor of each beam. By transforming the image to Fourier space, each component can be isolated individually, translated to the origin, and both the amplitude and the phase can be fully reconstructed by an inverse Fourier transformation \cite{Smits:2020}. To obtain normalized field of the probe beam, the second recording without atoms is made, from which the two complex phase factors can now be calculated and reconstructed, 
\begin{align}
        \tilde{E}_+ =& \frac{\left(E_p E_{R1}^*\right)_{\rm atoms}}{\left(E_{p} E_{R1}^*\right)_{\rm empty}} = e^{-i \phi_+} \\
        \tilde{E}_- =& \frac{\left(E_p E_{R2}^*\right)_{\rm atoms}}{\left(E_{p} E_{R2}^*\right)_{\rm empty}} = e^{-i \phi_-}.
\end{align}
Since the full field is obtained, one can calculate the field at different planes through beam propagation method (BMP) \cite{Thylen1983, Smits:2020}. 

\subsection{Sample Preparation}
The experiments are performed on a Bose-condensed gas of sodium atoms. Initially, the atoms undergo laser cooling and are confined within a cylindrically-symmetric magnetic trap (MT), achieving a pre-condensation temperature of approximately  $\sim 2 \ \mathrm{\mu K}$. After that, the atoms are transferred to an optical dipole trap (ODT) where the final stages of cooling are done. Such two-step cooling procedure ensures a maximal particle number, due to the enhanced cooling efficiency afforded by the magnetic trap as compared to the dipole trap. The ODT is formed by the 11 $\mathrm{\mu m}$ waist of a 1070 nm laser beam, with an initial optical power of 1.5 W. Just prior to the final stage of cooling, atoms are spin-flipped using a RF non-adiabatic passage between 2 MHz and 4 MHz at a constant magnetic field of 5.7 G. By varying the rate of the sweep, a desired population distribution between the three substates $m$ of the hyperfine level $F$, $\ket{F, m} = \ket{1, -1}$, $\ket{1, 0}$ and $\ket{1, 1}$ can be achieved, which was independently verified using Stern-Gerlach splitting technique. Finally, the trap beam power is lowered to $\sim 250$ mW, which produces an almost pure spinor condensate of about $2 \times 10^7$ atoms. The trap is cigar-shaped with a large aspect ratio, with trapping frequencies of $(\omega_r, \omega_a) = 2\pi \times (1260, 4.5)$ Hz. 

\begin{figure}[htbp]
\centering\includegraphics[width=0.48\textwidth]{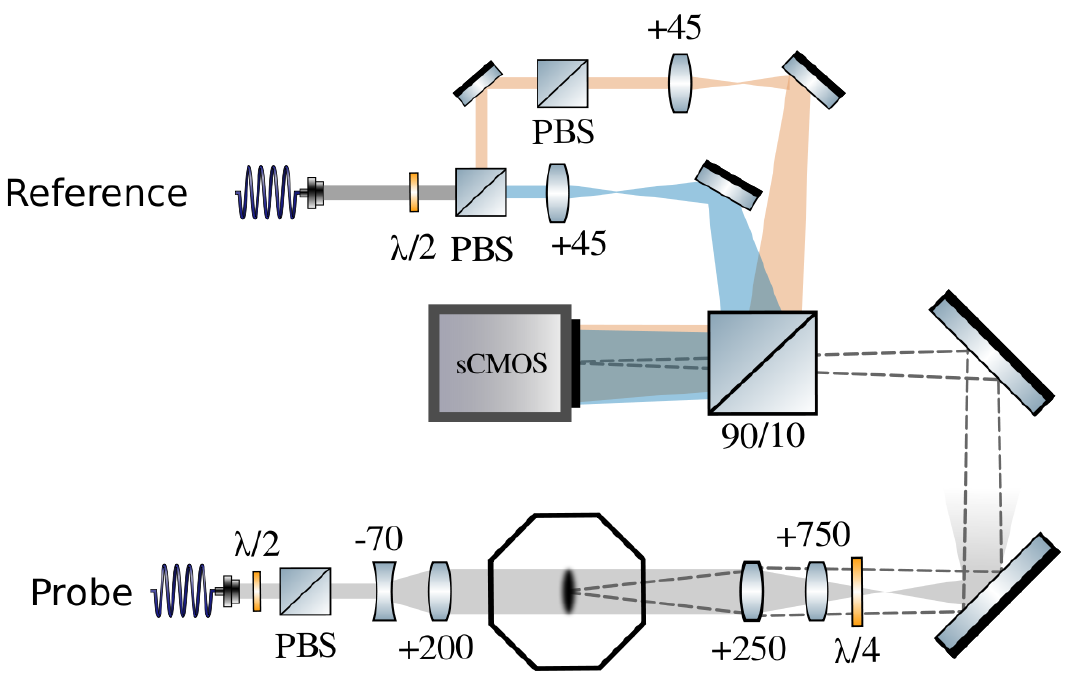}
\caption{Schematic representation of the setup, not to scale. The focal distance of lenses is given in millimeters. The octagon represents the vacuum chamber and is approximately 50 cm across. After outcoupling from the fiber a half-wave plate, denoted by $\lambda/2$, and a polarizing beam splitter (PBS) ensure proper polarization. The $\lambda/4$ plate converts the circular components of the probe light into perpendicular linear polarizations. The label sCMOS denotes the camera.}
\label{Fig:Setup}
\end{figure}

The atoms are imaged onto the camera using a linearly polarized probe beam (Fig. \ref{Fig:Setup}). On a separate optical table, the probe beam is split and both probe and reference beams are transported to the experimental setup via optical polarization maintaining fibers. The reference beam is further split up into two, orthogonally linearly polarized beams, which hit the camera and interfere with the probe beam under two different angles of few degrees. The probe and the two reference beams are detuned $-350$ MHz or approximately 36 atomic linewidths from the $F=1 \rightarrow F'=1$ $D_2$ transition ($\lambda = 589.16$ nm). The imaging frequency is in principle limited by the speed of the camera. Since the expected density and spin evolution of the ultracold cloud are roughly on the timescales set by the trapping frequencies, an imaging rate of  1 kHz is sufficient. The illumination pulses are 100 $\mathrm{\mu s}$ long, with an average intensity of 50 $\mathrm{\mu W/cm^2}$.  

\medskip

\section{Spin-Dependent Contrast and Spin Reconstruction} 

\subsection{Spin Imaging}\label{Sec:Spin_Density}

\begin{figure*}[htbp]
\centering\includegraphics[width=\textwidth]{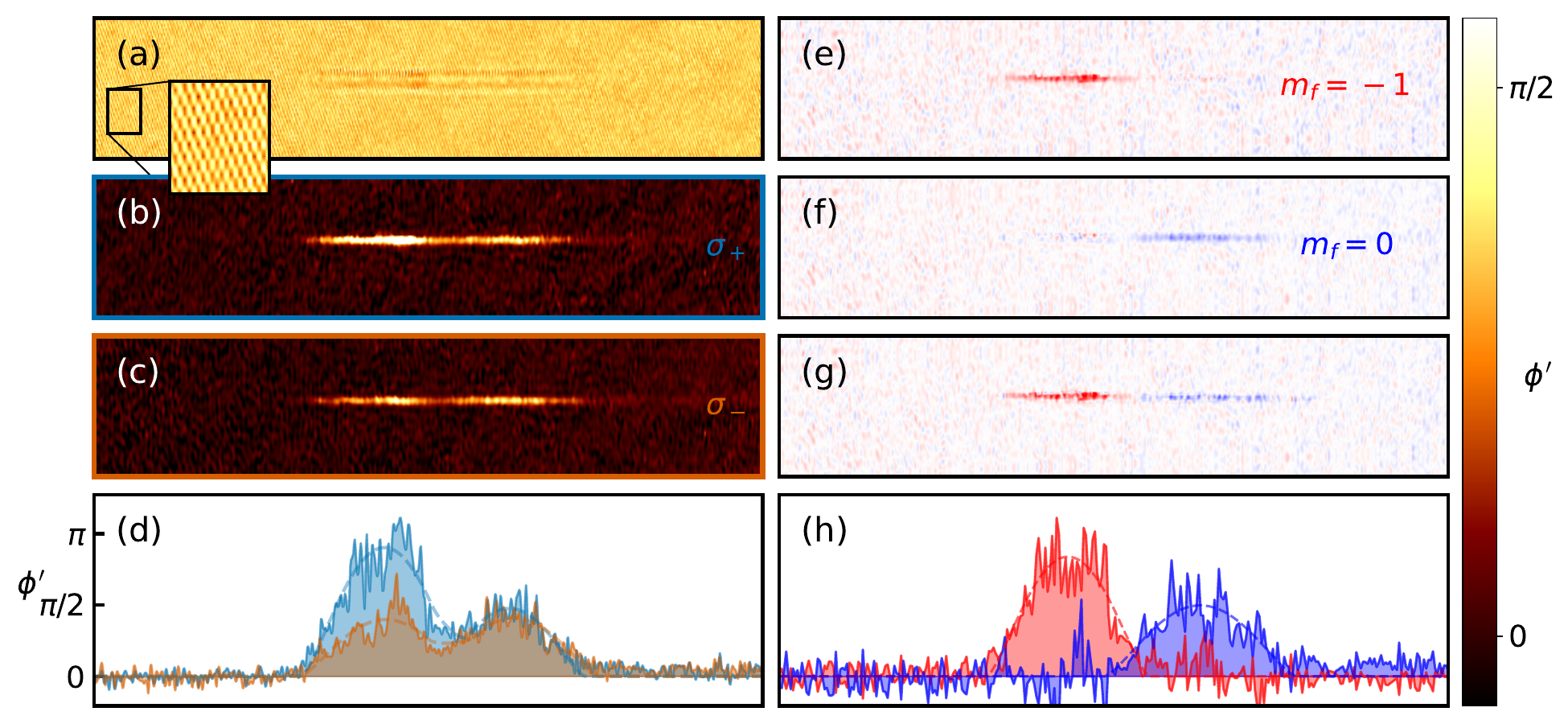}
\captionsetup{width=\textwidth}
\caption{Entire process of spin reconstruction, from a single raw image. \textbf{(a)} A double interference pattern is recorded on the camera, by interfering the circularly polarized probe with two orthogonally linearly polarized reference beams, under two different angles. Through Fourier analysis, the information about the phase delay experienced by \textbf{(b)} $\sigma_+$, and \textbf{(c)} $\sigma_-$ polarized light can be independently obtained (colorbar on the right). \textbf({d)} Linecuts through the central axis of the condensate confirm the spin-dependent contrast, with the atoms in $m = -1$ state imparting a larger phase delay on the $\sigma_+$ polarized part of the probe light (in blue) as opposed to $\sigma_-$ (in orange). The dashed lines indicate a fit of a sum of two Thomas-Fermi distributions. \textbf{(e) - (g)} Using the fit information, single spin components can be reconstructed based on relative contrasts between the two beams. All images \textbf{(a-c)} and \textbf{(e-g)} are 2450 $\mathrm{\mu m}$ in width and 150 $\mathrm{\mu m}$ in height, with condensate spanning roughly a 1000 $\mathrm{\mu m}$ in the axial and $10 \ \mathrm{\mu m}$ in radial direction.}
\label{Fig:MasterImage}
\end{figure*}

To demonstrate the working of SOAH, a condensate is prepared in a near-equal mixture of $m=0$ and $m=-1$ atoms. Since the method relies on a contrast between two beams, only two spin components can be reconstructed at a time, without making further assumptions. Following spinflipping, just prior to condensation, a magnetic field gradient in $z$-direction of 8.35 G/cm is applied for 25 ms, at the background bias of 5.7 G. This ensures that the condensate is in the polar regime, and will form domains \cite{Stenger:1998, TinLunHo:1998}. As predicted in \cite{Matuszewski:2008} and reported in \cite{Jimenez-Garcia:2019}, spontaneous domain formation can also be observed in the absence of any magnetic field gradient. After cooling, the quantization axis of the atoms is adiabatically rotated in direction of the probe beam through the use of the compensation coils, which are typically used to cancel out any stray magnetic fields. 

In order to separate the two components in the trap, a spin-dependent gradient of 8.35 G/cm is applied for a duration of 10 ms. Since the gradient of the dipole trap after the cooling is strongly reduced, this magnetic force is sufficient to spatially displace atoms in $m=-1$ state from the ones in $m = 0$ by $\approx$ 420 $\mathrm{\mu m}$. From the single raw image depicted in Fig. \ref{Fig:MasterImage}a, two images of phase shifts experienced by two orthogonal components of the probe beam are reconstructed (Fig. \ref{Fig:MasterImage}b, c). Linecuts along the central axis of both images (Fig. \ref{Fig:MasterImage}d) reveal a clear separation of the spin components. The $\sigma_+$ polarized beam shows a stronger interaction with the $m=-1$ atoms creating distinct phase delay contrast in the first peak, while both $\sigma_+$ and $\sigma_-$ beams equally affect the $m=0$ atoms, resulting in equal phase shifts for the second peak. The dashed lines represent a combination of two Thomas-Fermi models that fit the data, from which ratios between the peaks, and thus the contrast can be calculated. 

\begin{figure*}[htb]
\centering
\includegraphics[width=\textwidth]{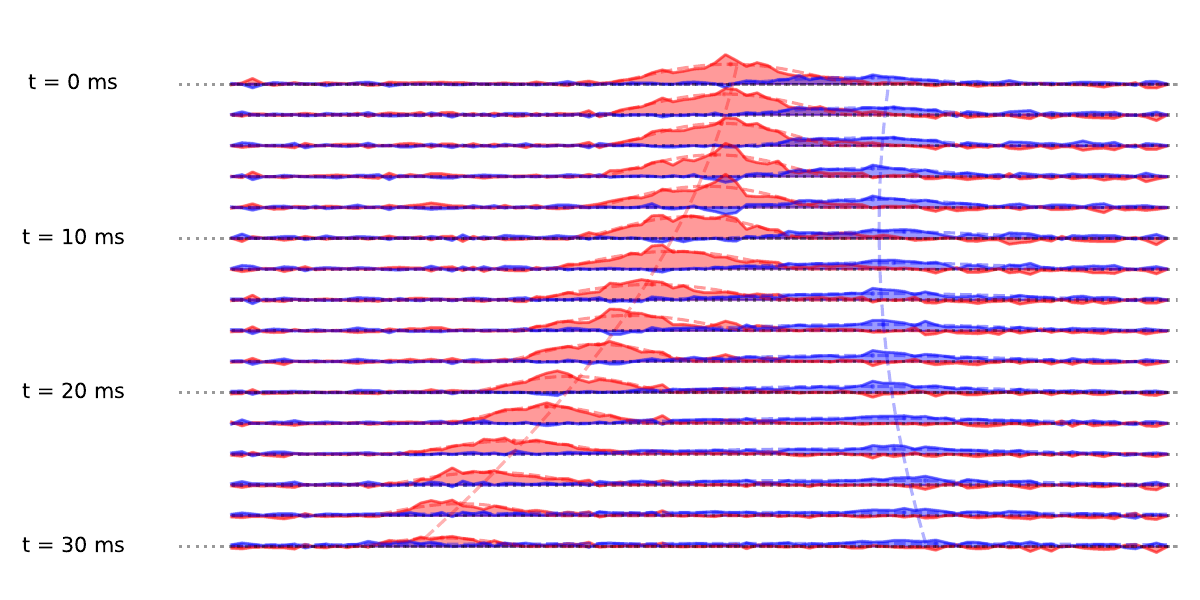}
\caption{A detailed examination of the dynamics within a multi-component spinor condensate under the influence of a spin-dependent force. The axial position of the different spin components as a function of time after the application of the spin-dependent force, with colors representing the different spin components. Each image is processed assuming the same contrast ratio between the two beams, which is calculated once the domains were far enough spatially separated.}
\label{Fig:Dynamics}
\end{figure*}

For the case of two spin components, the relationship between the accumulated phase of the two polarization's and spin density is simply described by
\begin{equation}
    \begin{pmatrix}
        \phi'_+ \\
        \phi'_-
    \end{pmatrix}
    =
    \begin{pmatrix}
        \sigma_{+1,0} &  \sigma_{+1,-1} \\
        \sigma_{-1,0} & \sigma_{-1,-1}
    \end{pmatrix}
    \begin{pmatrix}
        \rho^c_{0} \\
       \rho^c_{-1}
    \end{pmatrix},
\end{equation}
which can easily be inverted to yield spin density distribution. Experimentally found Thomas-Fermi amplitudes give optimal reconstruction ratios of $\phi'_+/\phi'_- = 1.02(7)$ to cancel out $m=0$ component, and $\phi'_+/\phi'_- = 0.54(3)$ to cancel out $m=-1$ component. Each reconstructed component is shown in Fig. \ref{Fig:MasterImage}e-f, with a fully spin resolved image shown in Fig. \ref{Fig:MasterImage}g, and the corresponding line cuts through the center shown in Fig. \ref{Fig:MasterImage}h. We note that in order to maximize the overlap of the two signals, the two reconstructed phase delays (Fig. \ref{Fig:MasterImage}b, c) must be slightly displaced relative to one another for roughly 1.5 $\mathrm{\mu m}$ in radial direction. At the moment, the cause of this shift is unknown, but we suspect the reason lies in the validity of the thin-lens approximation, and in the polarization specific birefringence effects.

The capabilities of this imaging technique for studying the dynamics of spinor condensates are further explored by tracking the position of the $m=0$ and $m=-1$ components \emph{in-situ} as illustrated in Fig. \ref{Fig:Dynamics}. The evolution of the single multi-component condensate is monitored after the application of a spin-dependent force. By reconstructing the axial positions of different spin components over time, analogous to the process presented in Fig. \ref{Fig:MasterImage}d, the centers of mass for each component are tracked, and the forces acting on the condensate are determined. A spin-dependent force of $F_s = 4.11 \times 10^{-26} \ \mathrm{N}$ and an acceleration of $a = 1.08 \ \mathrm{m / s^{2}}$ are measured, which is significantly lower than the expected acceleration of $a_{B} = 33 \ \mathrm{m / s^{2}}$ induced by the applied magnetic field gradient. The discrepancy between the two is attributed to the effects of the optical dipole trap and spin-drag forces. Enhanced precision in measurements could provide deeper insights into spin currents and spin resistivity, crucial aspects of spintronics.

\subsection{Magnetic field dependence} \label{Sec:MagFieldDependence}
The coefficients used to reconstruct each spin component in section \ref{Sec:Spin_Density} differ slightly from the theoretically predicted values in Table \ref{tab:CGs}. This is most likely caused by imperfect polarizations of the incoming beams, imperfections of the quarter wave plate used, and most notably, the fact that the magnetic field axis is not fully aligned with the wavevector of the probe light. In this section, these limitations are explored, and a more complete framework is derived, which allows for the reconstruction of not only spin composition, but also number density. 

\begin{figure*}[htbp]
\centering\includegraphics[width=\textwidth]{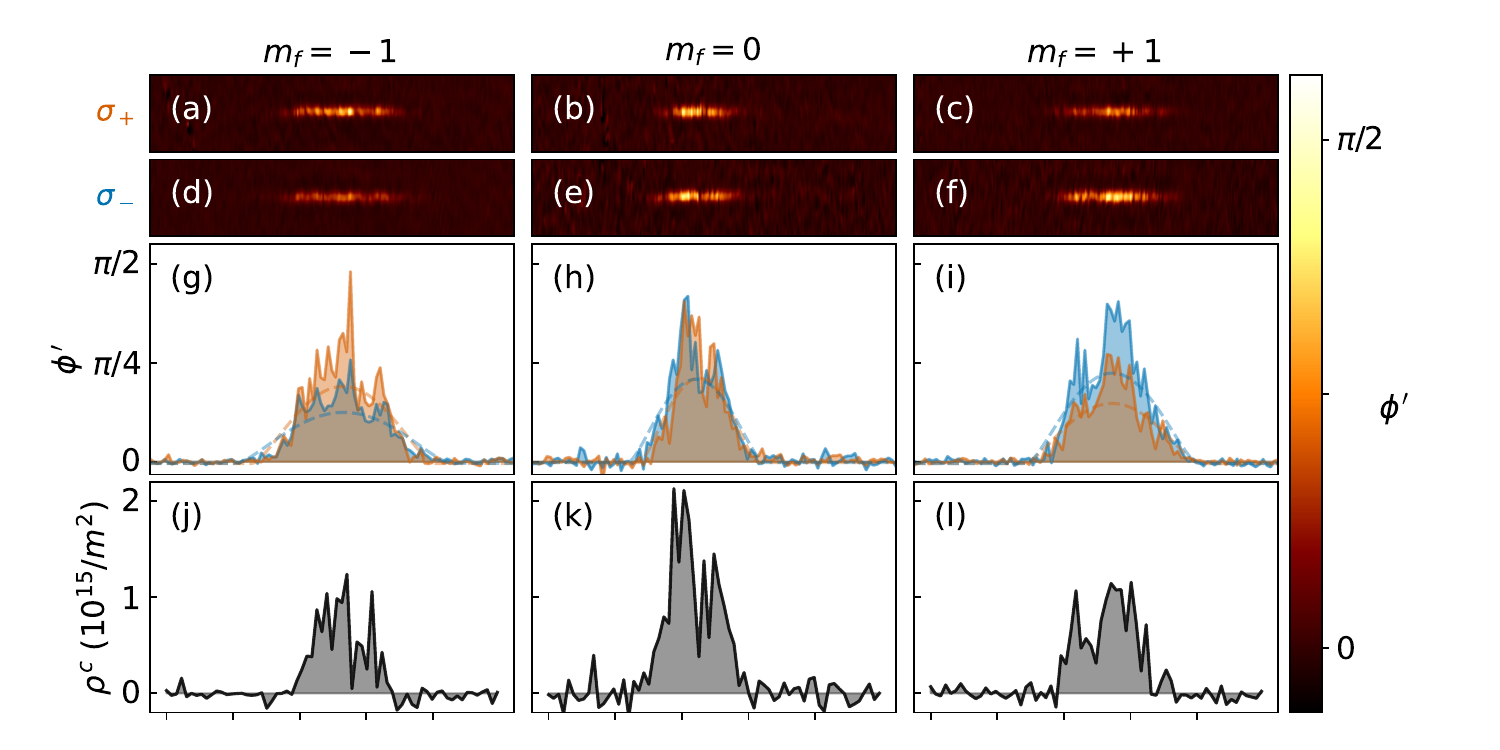}
\captionsetup{width=\textwidth}
\caption{Three different condensates prepared in different $m$ states imaged using SOAH. Since the magnetic field direction is known and is parallel to the light propagation axis, we can perfectly reconstruct all number densities, and calculate the corresponding particle numbers using theoretically known coefficients. Reconstructed phase delay for $m = -1, 0, +1$ states of the \textbf{(a-c)} $\sigma_+$ and of the \textbf{(d-f)} $\sigma_-$ component (colorbar on the right). Linecuts through the center of the density plots are indicated for both polarizations in \textbf{(g-i)}. The experimentally obtained ratios between the polarizations phase delays $\sigma_-/\sigma_+$ are $0.655$ for $m=-1$, $0.989$ for $m=0$ and $1.521$ for $m=+1$ component. The dimensions of the images are the same as the ones in Fig. \ref{Fig:MasterImage}. Ticks along the $x$-axis are spaced by 500 $\mu m$.}
\label{Fig:MIN_0_PLUS}
\end{figure*}

We start by creating a single component condensate at magnetic field known and aligned with the propagation direction of the light. The condensate is created in one of the three magnetic substates $m=-1, 0, +1$, and imaged using SOAH. The results for all three substates are shown in Fig. \ref{Fig:MIN_0_PLUS}. Using Eqs. (\ref{Eq:PhaseShiftTwoComponents}) and (\ref{Eq:M_Scattering_CS}) the column number density for each $m$ can be exactly calculated, from which properties such as the particle number and chemical potential can be extracted. Axial column density profiles along the centers of the BECs are shown in Fig. \ref{Fig:MIN_0_PLUS}j-l. For images depicted in Fig. \ref{Fig:MIN_0_PLUS}, the particle numbers are $N_{-1} = 3.17 \times 10^7$, $N_{0} = 1.79 \times 10^7$ and $N_{+1} = 1.04 \times 10^7$.

\begin{figure*}[t!]
    \centering
    \begin{subfigure}[t]{0.5\textwidth}
        \centering
        \includegraphics[width=\textwidth]{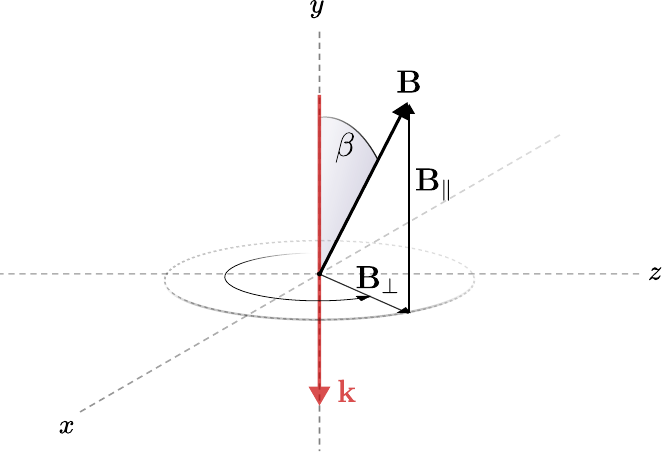} 
        \caption{} \label{Fig:Visualization}
    \end{subfigure}%
    ~ 
    \begin{subfigure}[t]{0.5\textwidth}
        \centering
        \includegraphics[width=\textwidth]{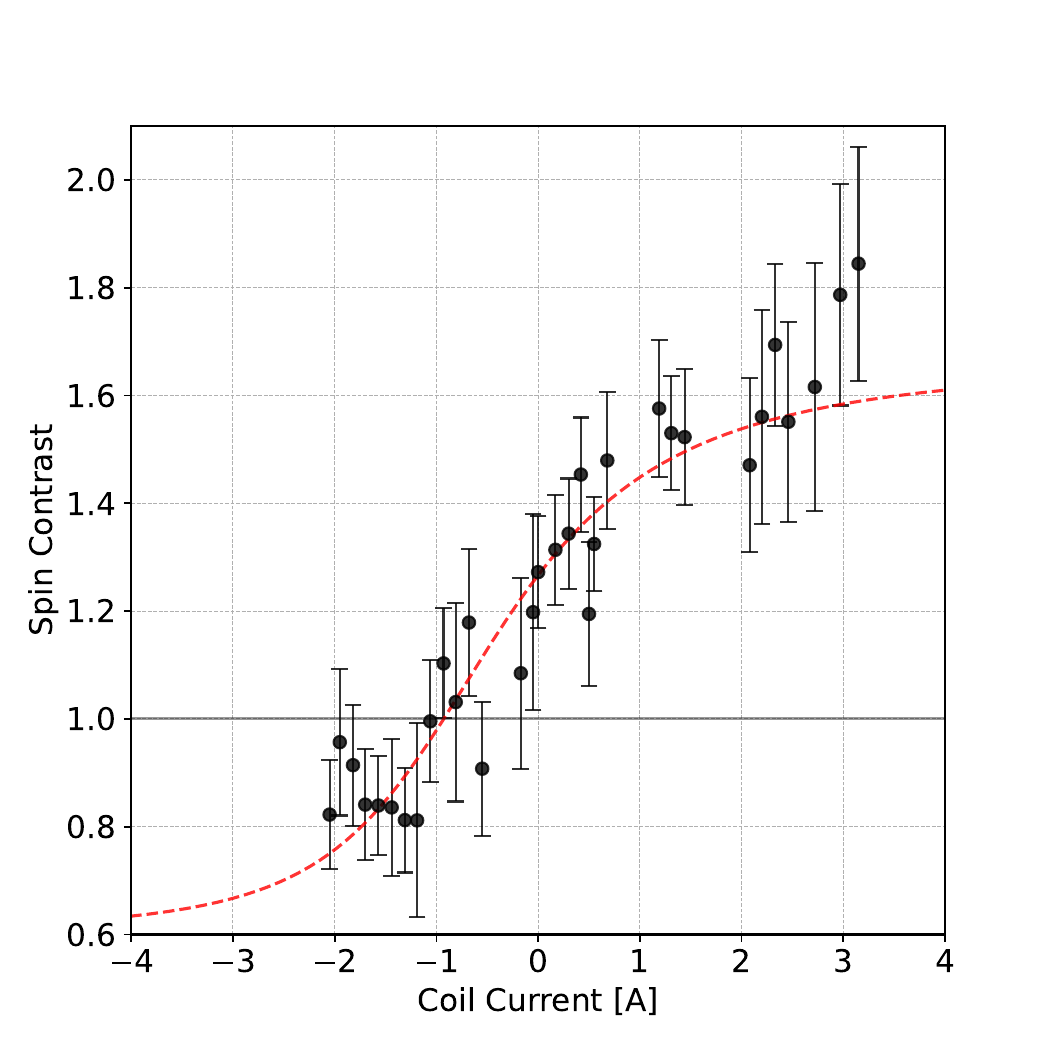}
        \caption{} \label{Fig:MagneticField}
    \end{subfigure}
    \captionsetup{width=\textwidth}
    \caption{\textbf{(a)} The angle $\beta$, defined by the direction of the quantization axis due to stray magnetic field components $\vec{B} = \vec{B}_\parallel + \vec{B}_\perp$, and the propagation of light $\vec{k}$. \textbf{(b)} The dependence of the spin induced contrast between $\sigma_+$ and $\sigma_-$ beams as a function of the current in the coils for atoms in $m = -1$ state. By varying the current through the compensation coils, the orientation of the resultant bias magnetic field vector can be changed, which leads to a difference in accumulated phase delay between two circularly polarized beams. The resulting distribution follows the relationship predicted by Eq. \ref{Eq:Contrast}, with angle $\beta$ given by Eq. (\ref{Eq:ARCTAN_fun}).}
\end{figure*}

In order to make SOAH applicable at a known magnetic field vector orientation, we investigate the impact of the orientation of the quantization axis on the phase delay contrast between the two circular polarizations, by varying the background static bias field.
The methodology relies on selectively varying the strength of the compensation field parallel to the direction of the beam propagation, $B_\parallel$. Assuming the presence of a static $B$-field in the plane perpendicular to the propagation of light, $\vec{B}_{\perp}$, this effectively rotates the resultant magnetic field vector $\vec{B} = \vec{B}_\parallel + \vec{B}_\perp$ with respect to the light propagation axis, $\vec{k}$, as shown in Fig. \ref{Fig:Visualization}. The angle $\beta$ is given by 
\begin{equation}\label{Eq:ARCTAN_fun}
    \beta = \frac{\pi}{2} - \tan^{-1} \Big( \frac{B_\parallel}{B_{\perp}} \Big).
\end{equation}

In order to account for the angle $\beta$ between the wavevector of the light and the magnetic field direction, one has to rotate the density matrix of the ground state over an angle $\beta$, as shown in \cite{Meppeling:2009}. The contrast between the two polarizations $q=\pm1$ for the $m=-1$ state changes from 5/3 for the situation described in Sec. 2.1 to the general case of $\beta \neq 0$:
\begin{equation}\label{Eq:Contrast}
    \frac{\phi_{-}}{\phi_{+}} = \frac{5\cos^4(\beta/2) + 2\sin^2(\beta) + 3\sin^4(\beta/2)}{3\cos^4(\beta/2) + 2\sin^2(\beta) + 5\sin^4(\beta/2)}. 
\end{equation}

To explore the impact of the angle $\beta$ on contrast, a condensate in the $m=-1$ spin state is imaged 50 times using SOAH, followed by phase delay reconstructions for each polarization. A Thomas-Fermi profile fitting is then performed on all 100 frames (50 per polarization), to calculate the average ratio of the phase delays. This procedure is repeated under various strengths of the parallel component of the magnetic field, $B_\parallel = \alpha I_\parallel + B_{\parallel,0}$, by adjusting the coil currents, $I_{\parallel}$. The factor $\alpha$ represents the coil's geometry-dependent linear conversion, and $B_{\parallel,0}$ represents a static background magnetic field. The data along with the fit are displayed in Fig. \ref{Fig:MagneticField}. Graphically, the horizontal offset of the fit corresponds to $B_{\parallel, 0}$, while the slope is related to the strength of the background magnetic field in the perpendicular plane, $B_\perp$. The two asymptotes converge to the theoretical values of the ratio between the $\sigma_+$ and $\sigma_-$ induced transitions in a spin-polarized atomic cloud, specifically $3/5$ and $5/3$.

Although the scope of investigation presented is limited to atoms anti-aligned with the magnetic field, $\ket{1, -1}$, by virtue of symmetry of atomic transitions, the $\sigma_+ / \sigma_-$ ratio is identical to the the ratio of $\sigma_- / \sigma_+$ for $\ket{1, +1}$ atoms. Furthermore, the inversion of the quantization axis for angles $\beta > \lvert \pi/2 \rvert$, which experimentally correspond to negative compensation currents, yields the opposite contrast. As the magnetic field is inverted, atoms in the state $\ket{1, -1}$, although remaining in the initial state relative to the lab frame, point in the opposite direction relative to the magnetic field - thus can be considered $\ket{1, +1}$ relative to the polarization of the probe. The same is true for atoms in the state $\ket{1, +1}$. 

The presented analysis technique enables the calibration necessary to determine the angle $\beta$ between the optical and magnetic axes, allowing for the implementation of SOAH at any non-zero $\beta$. Furthermore, the method serves as an experimental tool for measuring stray magnetic fields in cold atoms setups, similar to a recently proposed all-optical approach for measuring residual magnetic fields \cite{pomjaksilp2024}.

\section{Conclusion}
In conclusion, we introduce and validate the use of spin-dependent off-axis holography for non-invasive studies of spin-1 BECs, highlighting its significance as an advance in imaging technology and a catalyst for future research. Leveraging the interaction between light polarization and the spin states of the condensate, we achieve accurate reconstruction of the spin-density distribution and dynamics of the condensate \emph{in-situ} with minimal disturbance of the condensate. The insights gained from this technique are poised to deepen our understanding of spinor BECs, facilitating the development of new quantum technologies and contributing to the broader field of quantum information science.

\begin{acknowledgments}
The authors thank dr. Dries van Oosten for useful discussion, and P. Jurrius, J. B. Aans, D. Killian, and A. T. W. Driessen for technical support.
\end{acknowledgments}

\bibliography{Bibliography}
\end{document}